\documentclass[aps,prd,notitlepage,twocolumn,showpacs,nofootinbib,superscriptaddress]{revtex4-1}
\pdfoutput=1

\usepackage[utf8]{inputenc} 
\usepackage{graphicx}
\usepackage{tabularx}
\usepackage{amsmath}
\usepackage{amssymb}
\usepackage{float}
\usepackage{comment}
\usepackage{slashed}
\usepackage[normalem]{ulem}
\usepackage{caption}
\usepackage{subcaption}
\usepackage[usenames,dvipsnames]{color}
\usepackage[colorinlistoftodos]{todonotes}
\usepackage[colorlinks=true,citecolor=darkred,urlcolor=darkred, pdfborder={0 0 0}]{hyperref}
\usepackage[normalem]{ulem}
\usepackage[colorinlistoftodos]{todonotes}

\newcommand {\ignore}[1]{}

\usepackage{multirow}
\usepackage{subeqnarray}

\usepackage{scalefnt}

\definecolor{mightnightblue}{RGB}{25,25,112}
\definecolor{brown}{rgb}{0.59, 0.29, 0.0}
\definecolor{nicered}{rgb}{0.8,0.1,0.1}
\definecolor{nicegreen}{rgb}{0.1,0.5,0.1}
\definecolor{linkcolor}{rgb}{0,0,0.5}
\definecolor{darkred}{rgb}{0.6,0,0}

\def\gsim{\raise0.3ex\hbox{$\;>$\kern-0.75em\raise-1.1ex\hbox{$\sim\;$}}}
\def\lsim{\raise0.3ex\hbox{$\;<$\kern-0.75em\raise-1.1ex\hbox{$\sim\;$}}}

\usepackage{soul}

\def\21{$\mathrm{SU(2)_L \otimes U(1)_Y}$}

\def\cos{\rm {cos}}
\def\sin{\rm {sin}}

\definecolor{vdrgreen}{rgb}{0.0, 0.7, 0.0}

\newcommand{\AddrAHEP}{%
  AHEP Group, Institut de F\'{i}sica Corpuscular --
  CSIC/Universitat de Val\`{e}ncia, Parc Cient\'ific de Paterna.\\
 C/ Catedr\'atico Jos\'e Beltr\'an, 2 E-46980 Paterna (Valencia) - Spain}

\begin{document}



\title{\boldmath \color{BrickRed} Constraining nuclear physics parameters with current and future COHERENT data}

\author{D.K. Papoulias}\email{dipapou@ific.uv.es}\affiliation{\AddrAHEP}
\author{T.S. Kosmas}\email{hkosmas@uoi.gr}\affiliation{Division of Theoretical Physics, University of  Ioannina, GR 45110 Ioannina, Greece}
\author{R. Sahu}\email{rankasahu@gmail.com}\affiliation{National Institute of Science and Technology, Palur Hills, Berhampur-761008, Odisha, India}
\author{V.K.B. Kota}\email{vkbkota@prl.res.in}\affiliation{Physical Research Laboratory, Ahmedabad 380 009, India}
\author{M. Hota}\email{mihirhota@nist.edu}\affiliation{National Institute of Science and Technology, Palur Hills, Berhampur-761008, Odisha, India}

\begin{abstract}
  
Motivated by the recent observation of coherent elastic neutrino-nucleus
scattering (CE$\nu $NS) at the COHERENT experiment, our goal is to
explore its potential in probing important nuclear structure parameters.
We show that the recent COHERENT data offers unique opportunities to
investigate the neutron nuclear form factor. Our present calculations
are based on the deformed Shell Model (DSM) method which leads to a
better fit of the recent CE$\nu $NS data, as compared to known
phenomenological form factors such as the Helm-type, symmetrized Fermi
and Klein-Nystrand. The attainable sensitivities and the prospects of
improvement during the next phase of the COHERENT experiment are also
considered and analyzed in the framework of two upgrade scenarios. 

\end{abstract}

\maketitle
\noindent

\section{Introduction}%
\label{sec1}

The recent observation of coherent elastic neutrino nucleus scattering
(CE$\nu $NS) events at the Spallation Neutron Source (SNS) by the
COHERENT experiment~\cite{Akimov:2017ade,Akimov:2018vzs}, has
opened up new opportunities to probe physics in theories within and
beyond the Standard Model (SM) of electroweak interactions. The COHERENT
program is aiming to investigate several important physical phenomena
through low-energy precision measurements. The first CE$\nu $NS
observation has triggered the theoretical challenges required to
interpret neutrino-nuclear responses~\cite{Ejiri:2019ezh} in the
context of new physics models~\cite{Kosmas:2017tsq}.

Recently, several studies were conducted in trying to analyze and
interpret the COHERENT data, in order to examine possible deviations
from the SM predictions that may point to new
physics~\cite{Barranco:2005yy,Scholberg:2005qs}. These searches
address non-standard interactions
(NSIs)~\cite{Liao:2017uzy,Dent:2017mpr,AristizabalSierra:2017joc,Denton:2018xmq},
electromagnetic (EM)
properties~\cite{Kosmas:2015sqa,Billard:2018jnl,Cadeddu:2018dux},
sterile
neutrinos~\cite{Kosmas:2017zbh,Canas:2017umu,Blanco:2019vyp}, novel
mediators~\cite{Dent:2016wcr,Farzan:2018gtr,Abdullah:2018ykz,Brdar:2018qqj},
CP-violation~\cite{Miranda:2019wdy,AristizabalSierra:2019ufd} and
implications to dark
matter~\cite{Ge:2017mcq,Coloma:2017ncl,Gonzalez-Garcia:2018dep}.
Potential contributions due to neutrino-nucleus scattering at direct dark
matter detection detectors have been
explored~\cite{Ng:2017aur,Papoulias:2018uzy,Boehm:2018sux,Dutta:2019oaj},
while the CE$\nu $NS cross section has been also revisited
within~\cite{Bednyakov:2018mjd} and beyond the
SM~\cite{Lindner:2016wff,AristizabalSierra:2018eqm,Miranda:2019skf}.

The nuclear form factor related to weak interactions plays a dominant
role in the accurate description of neutrino-
matter interactions~\cite{Payne:2019wvy}
motivating further the necessity of revisiting the relevant nuclear parameters (see 
Refs.~\cite{Cadeddu:2017etk,Yang:2019pbx}). 
While neutrinos are a valuable tool for deep sky
investigations~\cite{Smponias:2015nua}, nuclear 
 parameters such as the neutron skin can be 
crucial for understanding neutron 
star dynamics~\cite{Fattoyev:2017jql}. In
this work we explore how such nuclear parameters 
can be probed at CE$\nu$NS
experiments. For realistic nuclear structure calculations, we employ the
deformed shell model (DSM) based on Hartree-Fock (HF) deformed intrinsic
states with angular momentum projection and band
mixing~\cite{ks-book}. The DSM has been previously applied for
describing nuclear spectroscopic
properties~\cite{ks-book,Sahu:2013yna,Sahu:2014nga}, exotic
processes such as $\mu \to e$ conversion in
nuclei~\cite{Kosmas:2003xr} and WIMP-nucleus
scattering~\cite{Sahu:2017czz}.

The conventional neutrino-processes are theoretically
well-studied~\cite{Papoulias:2015vxa,Pirinen:2018gsd}, while the
recent CE$\nu $NS observation motivates precision tests of the SM at low
energies~\cite{Kosmas:2015vsa}. It has been shown that a
competitive determination of the weak-mixing angle is
possible~\cite{Canas:2018rng}, while CE$\nu $NS also highlights
a novel avenue for probing the neutron nuclear form
factor~\cite{Patton:2012jr,Cadeddu:2017etk,Ciuffoli:2018qem}.
During its phase I, the COHERENT collaboration achieved a high
experimental sensitivity and a low detector threshold which led to the
first observation of CE$\nu $NS while also intends to enhance its future
program with a multitarget strategy~\cite{Akimov:2018ghi}. Apart
from the next phase of COHERENT, other experiments are planned to
operate with reactor neutrinos like the
TEXONO~\cite{Wong:2010zzc},
CONNIE~\cite{Aguilar-Arevalo:2016qen},
MINER~\cite{Agnolet:2016zir}, $\nu $GEN~\cite{Belov:2015ufh},
CONUS~\cite{conus}, Ricochet~\cite{Billard:2016giu} and
NU-CLEUS~\cite{Strauss:2017cuu}, further motivating the
present work.

Muon spectroscopy~\cite{Fricke:1995zz} and atomic parity violating
(APV) electron scattering data~\cite{Angeli:2013epw} from the PREX
experiment~\cite{Abrahamyan:2012gp} has been employed as a
powerful tool to measure the spatial distributions of neutrons in
nuclei~\cite{Horowitz:2012tj,Orrigo:2016wgu,Cadeddu:2018izq}. Our
paper focuses on the open issues related to constraining the nuclear
physics parameters~\cite{Huang:2019ene,AristizabalSierra:2019zmy}
entering the description of the weak neutral current vector and axial
vector properties, such as ground state properties mostly related to the
dominance of neutrons participating in the materials of rare-events
detectors~\cite{Lewin:1995rx}. On the basis of our nuclear DSM
calculations and the COHERENT data, we will make an attempt to extract
constraints on the nuclear form factors in the
Helm~\cite{Helm:1956zz}, symmetrized
Fermi~\cite{Sprung_1997} and
Klein-Nystrand~\cite{Klein:1999qj} approach, as well as to explore
the neutron radial moments~\cite{Piekarewicz:2016vbn}.

The paper has been organized as follows: in Sect.~\ref{sec:cevns} we
present the relevant formalism to accurately simulate the COHERENT data,
while in Sect.~\ref{sec:form-fac} we introduce the DSM method and
discuss the various form factor parametrizations considered.
Sect.~\ref{sec:result} presents the main outcomes of this work and
finally in Sect.~\ref{sec:conclusion} the main conclusions are discussed.

\section{CE$\nu $NS within deformed shell model calculations}%
\label{sec:cevns}

%
\begin{table*}[ht]
\centering
\begin{tabular}{ccccccccccc}
\hline
Nucleus & $A$ & $Z$ & $J^{\pi }$ & $<l_{p}>$ & $<S_{p}>$ & $<l_{n}>$ & $<S_{n}>$ & $\mu $ (nm) & Exp (nm) & $b \, \, [\mathrm{fm^{-1}}]$ \\
\hline
I & 127 & 53 & $5/2^{+}$ & 2.395 & $-0.211$ & 0.313 & 0.002 & 1.207 & 2.813 & 2.09\\
Cs & 133 & 55 & $7/2^{+}$ & 3.40 & $ -0.339$ & 0.49 & $-0.048$ & 1.69 & 2.582 & 2.11\\
\hline
\end{tabular}
\caption{The calculated magnetic moments and their decomposition into orbital
and spin parts for the ground states of $^{127}$I and $^{133}$Cs. The magnetic
moments given in column 9 are obtained by multiplying the entries in columns 5-8
with the bare gyromagnetic ratios (in nm units) $g^{p}_{l}=1$, $g^{n}_{l}=0$,
$g^{p}_{s}=5.586$ and $g^{n}_{s}=-3.826$ and then summing. Shown in the table
are also the ground state $J^{\pi }$ and the harmonic oscillator size parameter
$b$ employed in the calculations. The experimental data are taken from
Ref.~\cite{nndc}.}
\label{tab-1}
\end{table*}

Within the framework of the SM, the CE$\nu $NS differential cross
section with respect to the nuclear recoil energy $T_{A}$ is written
as~\cite{Kosmas:2017tsq,Papoulias:2018uzy}
%
\begin{equation}
\frac{d \sigma }{dT_{A}} = \frac{G_{F}^{2} m_{A}}{\pi } \left [G_{V}
^{2} \left (1 - \frac{m_{A} T_{A}}{2 E_{\nu }^{2}}\right ) + G_{A}
^{2} \left (1 + \frac{m_{A} T_{A}}{2 E_{\nu }^{2}}\right ) \right ]
\, ,
\end{equation}
where $G_{F}$ is the Fermi coupling constant, $E_{\nu }$ is the neutrino
energy and $m_{A}$ the nuclear mass of the target $(A,Z)$, with $Z$
protons and $N=A-Z$ neutrons ($A$ is the mass number). The vector and
axial vector weak charges $G_{V}$ and $G_{A}$, depend on the momentum
variation of the proton and neutron nuclear form factors $F_{p}(Q^{2})$
and $F_{n}(Q^{2})$, as~\cite{Bednyakov:2018mjd}
%
\begin{equation}
\begin{aligned}
G_{V}(Q) =
& \left [ g^{V}_{p} Z F_{p}(Q^{2}) + g^{V}_{n} N F_{n}(Q
^{2}) \right ] \, , \\
G_{A}(Q) =& \left [ g^{A}_{p} (\delta Z) F_{p}(Q
^{2}) + g^{A}_{n} (\delta N) F_{p}(Q^{2}) \right ] \, ,
\end{aligned}
\end{equation}
with the vector couplings for protons and neutrons taken as
$g_{p}^{V}=1/2-2 \sin ^{2} \theta _{W}$ and $g_{n}^{V}=-1/2$ respectively,
and the weak mixing angle $\theta _{W}$ fixed to the PDG value
$\sin ^{2} \theta _{W}=0.2312$~\cite{Beringer:1900zz}. The
corresponding axial vector couplings for protons and neutrons are
defined as $g_{p}^{A}=1/2$ and $g_{n}^{A}=-1/2$, while $(\delta Z)=Z
_{+} - Z_{-}$ and $(\delta N)=N_{+} - N_{-}$, where the $+$ or $-$ sign
accounts for the total number of protons or neutrons with spin up and
down, respectively~\cite{Barranco:2005yy}. Note that the
$g_{A}$ couplings are quenched for charged-current processes (see
Refs.~\cite{Ejiri:2019ezh,Suhonen:2017rjf}).

The COHERENT experiment has made the first ever observation of CE$
\nu $NS with a CsI[Na] detector of mass $m_{\mathrm{det}}=14.57$~kg
exposed to neutrino emissions from the $\pi$-DAR source at a distance
of $L=19.3$~m, for a period of $t_{\mathrm{run}}=308.1$~days. To
adequately simulate the recent COHERENT data we consider the total cross
section as the sum of the individual cross sections by taking also into
account the stoichiometric ratio $\eta $ of the corresponding atom. For
a given neutrino flavor $\alpha $ and isotope $x$, the number of CE$
\nu $NS events reads~\cite{Kosmas:2017tsq}
%
\begin{equation}
\begin{aligned}
N_{\mathrm{theor}} = \sum _{\nu _{\alpha }} \sum _{x = \mathrm{Cs, I}} &
\mathcal{F}_{x}
 \int _{E_{\nu }^{\mathrm{min}}}^{E_{\nu }^{
\mathrm{max}}} \lambda _{\nu _{\alpha }}(E_{\nu }) dE_{\nu }\\ \times &
\int _{T_{A}^{\mathrm{min}}}^{T_{A}^{\mathrm{max}}} \mathcal{ A } ( T
_{A} ) \frac{{d \sigma }_{x}}{dT_{A}}(E_{\nu }, T_{A}) dT_{A} \, ,
\end{aligned}
\label{eq:events}
\end{equation}
where
\begin{equation}
\mathcal{F}_{x} = t_{\mathrm{run}} N_{\mathrm{targ}}^{x} \Phi _{\nu }
\, .
\label{eq:F}
\end{equation}
The neutrino flux is $\Phi _{\nu } = r \mathcal{N}_{\mathrm{POT}}/4
\pi L^{2}$, with $r=0.08$ representing the number of neutrinos per
flavor produced for each proton on target (POT), where $\mathcal{N}
_{\mathrm{POT}} = N_{\mathrm{POT}}/t_{\mathrm{run}}$ with $N_{
\mathrm{POT}}=1.76 \times 10^{23}$. Our calculations consider the Geant4
SNS neutrino spectrum taken from the upper panel of Fig.~S2 shown in
Ref.~\cite{Akimov:2017ade}. Here, the various flavor components
$\nu _{\alpha }=\{ \nu _{e}, \nu _{\mu }, \bar{\nu }_{\mu }\}$ of the SNS
neutrino spectrum, including also the monochromatic
$E_{\nu _{\mu }}=29.9$~MeV prompt beam from pion decay at rest, are
denoted as $\lambda _{\nu _{\alpha }}(E_{\nu })$, while for each isotope
$x = \mathrm{Cs, I}$, the number of target nuclei is expressed in terms
of Avogadro's number $N_{A}$ and the detector mass
%
\begin{equation}
N_{\mathrm{targ}}^{x} = \frac{m_{\mathrm{det}} \eta _{x}}{\sum _{x} A
_{x} \eta _{x}} N_{A} \, .
\end{equation}
We furthermore stress that contributions to event rate from the sodium
dopant are of the order $10^{-5}$--$10^{-4}$ and can be safely
ignored~\cite{Collar:2014lya}.

The recent observation of the CE$\nu $NS signal at COHERENT experiment
was based on photoelectron (PE) measurements. To translate the nuclear
recoil energy in terms of the number of PE, $n_{\text{PE}}$, we adopt
the relation~\cite{Akimov:2017ade}
%
\begin{equation}
n_{\mathrm{PE}} = 1.17 \frac{T_{A}}{(\mathrm{keV})}\, .
\end{equation}
In Eq.~(\ref{eq:events}), the photoelectron dependence of the detector
efficiency $\mathcal{A}(x)$ is given by the
expression~\cite{Akimov:2018vzs}
%
\begin{equation}
\mathcal{ A } ( x ) = \frac{ k_{ 1 } }{ 1 + e ^{ - k_{ 2 } \left ( x -
x_{ 0 } \right ) } } \Theta (x) \, ,
\end{equation}
with parameters $k_{1}= 0.6655$, $k_{2}= 0.4942$, $x_{0}= 10.8507$ and
$\Theta (x)$ being the Heaviside function, defined as
%
\begin{equation}
\Theta ( x ) = \left \{
\begin{array}{ l@{\quad } l }
{ 0 } & { x < 5 }
\\
{ 0.5 } & { 5 \leq x < 6 }
\\
{ 1 } & { x \geq 6 \, .}
\end{array}
\right .
\end{equation}

\section{Evaluation of the nuclear form factors}%
\label{sec:form-fac}

In CE$\nu $NS and direct dark matter detection searches, to account for
the finite nuclear size, the nuclear form factor is defined as the
Fourier transform of the nuclear charge density
distribution~\cite{Papoulias:2015vxa}
%
\begin{equation}
F_{n,p}(Q^{2}) = \frac{1}{N_{a}} \int \rho _{p,n}(\vec{r}) \, e^{i
\vec{Q} \cdot \vec{r}} \, d^{3} \vec{r}, \quad N_{a}={Z,N} \, ,
\end{equation}
with $F_{p} \neq F_{n}$. Following a model independent approach, the
nuclear form factor can be expanded in terms of even moments of the
charge density distribution~\cite{Patton:2012jr}
%
\begin{equation}
\begin{aligned}
F_{p,n} (Q^{2}) \approx 1 - \frac{Q^{2}}{3!} \langle R_{p,n}^{2}
\rangle + \frac{Q^{4}}{5!} \langle R_{p,n}^{4} \rangle -
\frac{Q^{6}}{7!} \langle R_{p,n}^{6} \rangle + \cdots \, ,
\end{aligned}
\label{eq:form-factor-expansion}
\end{equation}
with the $k$-th radial moment defined as
%
\begin{equation}
\langle R_{p,n}^{k} \rangle =\frac{\int \rho _{p,n}(\vec{r}) \, r^{k}
\, d^{3} \vec{r}}{\int \rho _{p,n}(\vec{r}) \, d^{3} \vec{r}} \, .
\end{equation}

From experimental physics perspectives, it is feasible to measure only
the proton charge density distribution with high precision from electron
scattering data~\cite{Angeli:2013epw}. For this reason, numerous
studies rely on the approximation $\rho _{p} = \rho _{n}$ and thus assume
$F_{p} = F_{n}$. On the theoretical side, both the proton and neutron
nuclear form factors can be treated separately, within the context of
advanced nuclear physics methods such as, the large-scale
Shell-Model~\cite{Kortelainen:2006rd,Toivanen:2009zza}, the
Quasiparticle Random Phase Approximation
(QRPA)~\cite{Papoulias:2013gha}, Microscopic Quasiparticle Phonon
Model (MQPM)~\cite{Pirinen:2018gsd} and the method of DSM
calculations~\cite{Papoulias:2018uzy}. In the present work we
employ the latter method. Our primary goal is to extract crucial
information on the nuclear parameters entering the various form factor
approaches from the recent data of the COHERENT experiment, relying on
the various definitions of the nuclear form factor that we consider in
the present study.

In the concept of DSM, for the calculation of the form factors relevant
to the COHERENT detector materials $^{127}$I and $^{133}$Cs, we have
adopted an effective interaction recently developed in
Ref.~\cite{Coraggio:2017bqn} employing a model space consisting
of the spherical orbitals $0g_{7/2}$, $1d_{5/2}$, $1d_{3/2}$,
$2s_{1/2}$ and $0h_{11/2}$ with the closed core $^{100}$Sn. The
effective interaction is obtained by renormalizing the CD-Bonn
potential. The single particle energies for the five orbitals are taken
to be 0.0, 0.4, 1.4, 1.3 and 1.6 MeV for protons and 0.0, 0.7, 2.1, 1.9
and 3.0 MeV for neutrons. We first perform an axially symmetric HF
calculation and obtain the lowest intrinsic solution using the above
effective interaction for each of the above nuclei. Then, excited
intrinsic states are obtained by making particle-hole excitations over
the lowest intrinsic states. At the final step, we perform angular
momentum projection and band mixing and obtain the nuclear wave
functions which are used for calculating different properties of these
nuclei. We stress that including more orbits requires a new effective
interaction that is beyond the scope of the present paper.

We have considered six intrinsic configurations for $^{127}$I and three
intrinsic configurations for $^{133}$Cs. These intrinsic states are
found to be sufficient to produce most of the important properties of
these isotopes (complete details will be reported elsewhere). In
Table~\ref{tab-1}, we tabulate the most important observables and
outcomes of the nuclear structure calculations from DSM in the present
work. Specifically, the observables include the magnetic moments of the
two nuclei considered and the contribution of protons and neutrons to
the orbital and spin parts giving better physical insight. Magnetic
moments and spectroscopic properties of the two nuclei are calculated
to check the reliability of the nuclear wave functions generated by DSM.

Besides realistic nuclear structure calculations within DSM, a rather
reliable description of the nuclear form factor is the known as Helm
approximation. The latter relies on the convolution of two nucleonic
densities, one being a uniform density with cut-off radius,
$R_{0}$, (namely box or diffraction radius) characterizing the interior
density and a second one that is associated with a Gaussian falloff in
terms of the surface thickness, $s$. In the Helm approximation the form
factor is expressed in analytical form as~\cite{Helm:1956zz}
%
\begin{equation}
F_{\text{Helm}}(Q^{2}) = 3 \frac{j_{1}(Q R_{0})}{q R_{0}}\, e^{-(Q
s)^{2}/2} \, ,
\label{eq:helm}
\end{equation}
where $j_{1}(x)$ denotes the 1st-order spherical Bessel function. The
first three moments can be analytically expressed
as~\cite{Piekarewicz:2016vbn}
%
\begin{equation}
\begin{aligned}
\left \langle R^{2}_{n} \right \rangle & = \frac{ 3 }{ 5 } R _{ 0 }
^{ 2 } + 3 s ^{ 2 } \\
\left \langle R^{4}_{n} \right \rangle & = \frac{
3 }{ 7 } R _{ 0 } ^{ 4 } + 6 R _{ 0 } ^{ 2 } s ^{ 2 } + 15 s ^{ 4 }
\\
\left \langle R^{6}_{n} \right \rangle & = \frac{ 1 }{ 3 } R _{ 0 }
^{ 6 } + 9 R _{ 0 } ^{ 4 } s ^{ 2 } + 63 R _{ 0 } ^{ 2 } s ^{ 4 } + 105
s ^{ 6 } \, .
\end{aligned}
\end{equation}
%
\begin{figure*}[t]
\centering
\includegraphics[width=0.9\linewidth]{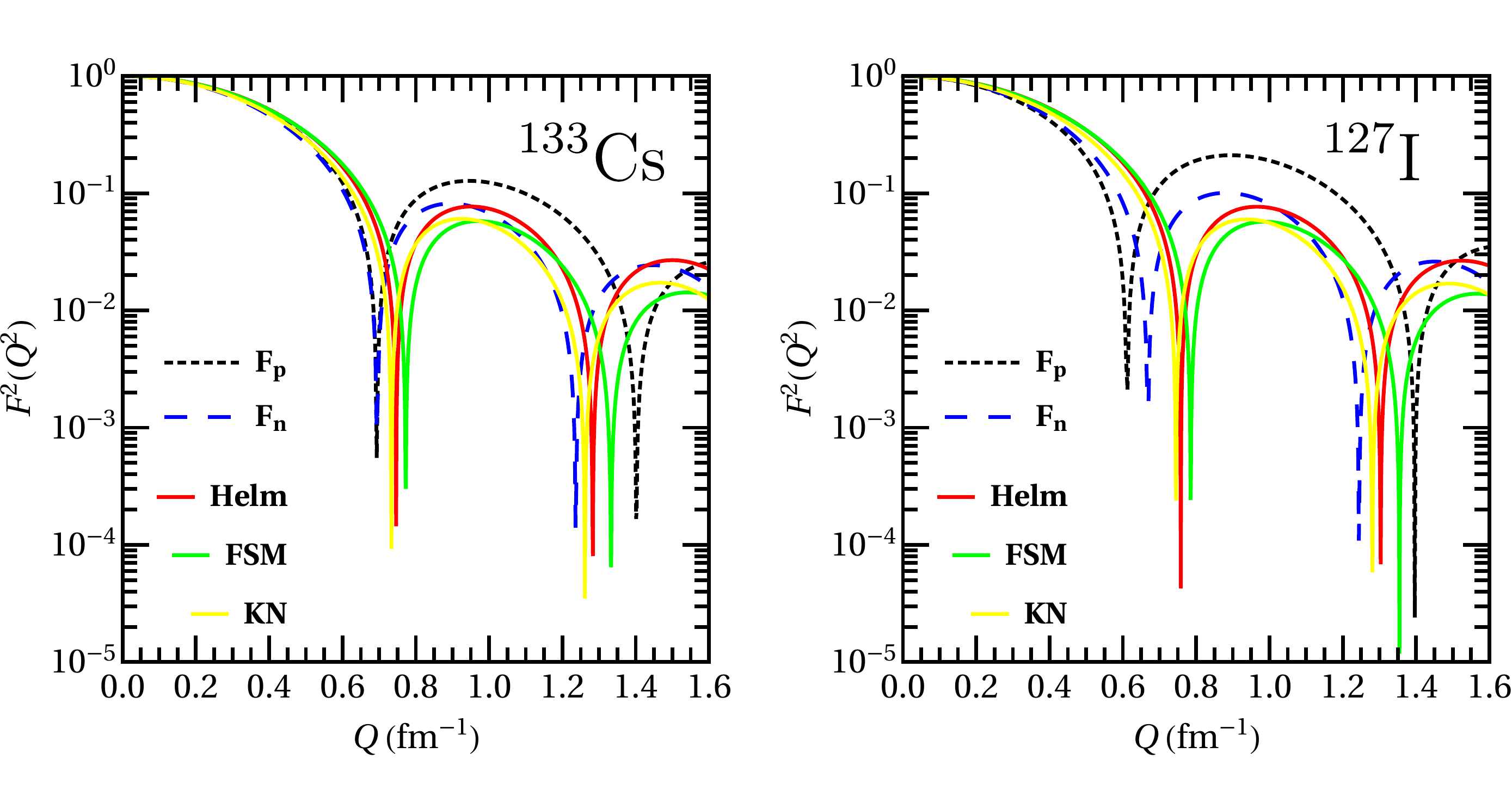}
\caption{Proton and neutron weak nuclear form factors of $^{133}$Cs (left) and
$^{127}$I (right) nuclei as a function of the momentum transfer
$Q(\text{fm}^{-1})$, calculated with DSM and compared with Helm, SF and KN form
factors.}
\label{fig:form-factor}
\end{figure*}
Following Ref.~\cite{Lewin:1995rx} we fix an \emph{ad-hoc} value
$s=0.9$, obtained by fitting to muon spectroscopy
data~\cite{Fricke:1995zz}. The latter has the advantage of
improving the matching between the Helm and the symmetrized Fermi (SF)
form factor that is discussed below. Adopting a conventional Fermi
(Woods-Saxon) charge density distribution, the SF form factor is written
in terms of two parameters $(c,a)$ in analytical form,
as~\cite{Sprung_1997}
%
\begin{equation}
\begin{aligned}
F_{\text{SF}} \left ( Q ^{ 2 } \right ) =
& \frac{ 3 }{ Q c \left [ (
Q c ) ^{ 2 } + ( \pi Q a ) ^{ 2 } \right ] } \left [ \frac{ \pi Q a }{
\sinh ( \pi Q a ) } \right ] \\ & \times \left [ \frac{ \pi Q a
\sin ( Q c ) }{ \tanh ( \pi Q a ) } - Q c \cos ( Q c ) \right ] \, ,
\end{aligned}
\end{equation}
with
%
\begin{equation}
c = 1.23 A^{1/3} - 0.60 \, \text{(fm)}, \quad a=0.52 \, \text{(fm)}
\, ,
\label{eq:SF-vals}
\end{equation}
representing the half density radius and the diffuseness respectively.
The surface thickness in this case is quantified through the relation
$t = 4a \ln 3 $~\cite{Cadeddu:2017etk}. In
Ref.~\cite{Piekarewicz:2016vbn} the first three moments entering
Eq.~(\ref{eq:form-factor-expansion}) are expressed in analytical form,
for the case of the Fermi symmetrized form factor, as
%
\begin{equation}
\begin{aligned}
\left \langle R^{2}_{n} \right \rangle & = \frac{ 3 }{ 5 } c ^{ 2 } +
\frac{ 7 }{ 5 } ( \pi a ) ^{ 2 } \\ \left \langle R^{4}_{n}
\right \rangle & = \frac{ 3 }{ 7 } c ^{ 4 } + \frac{ 18 }{ 7 } (
\pi a ) ^{ 2 } c ^{ 2 } + \frac{ 31 }{ 7 } ( \pi a ) ^{ 4 } \\ \left \langle R
^{6}_{n} \right \rangle & = \frac{ 1 }{ 3 } c ^{ 6 } + \frac{ 11 }{ 3
} ( \pi a ) ^{ 2 } c ^{ 4 } + \frac{ 239 }{ 15 } ( \pi a ) ^{ 4 } c
^{ 2 } + \frac{ 127 }{ 5 } ( \pi a ) ^{ 6 } \, .
\end{aligned}
\end{equation}

The COHERENT collaboration, has adopted the Klein-Nystrand (KN) form
factor which follows from the convolution of a Yukawa potential with
range $a_{k} = 0.7$ fm over a Woods-Saxon distribution, approximated as
a hard sphere with radius $R_{A}$. The resulting form factor
reads~\cite{Klein:1999qj}
%
\begin{equation}
F_{\text{KN}} = 3 \frac{j_{1}(Q R_{A})}{Q R_{A}} \left [ 1 + (Q a_{k}
)^{2} \right ]^{-1} \, ,
\end{equation}
whereas the corresponding root mean square (rms) radius becomes
%
\begin{equation}
\langle R^{2} \rangle _{\text{KN}} = 3/5 R_{A}^{2} + 6 a_{k}^{2} \, .
\end{equation}

The form factor evaluated with DSM calculations is illustrated in
 Fig.~\ref{fig:form-factor} and is compared with the Helm, SF and KN
parametrizations. As can be seen, in general, $F_{p} = F_{n}$ is not
always a good approximation since minima and maxima of $F_{p}$ and
$F_{n}$ occur at different values of the momentum transfer.


\section{Results and discussion}%
\label{sec:result}

\begin{figure*}
\includegraphics[width=0.9\textwidth]{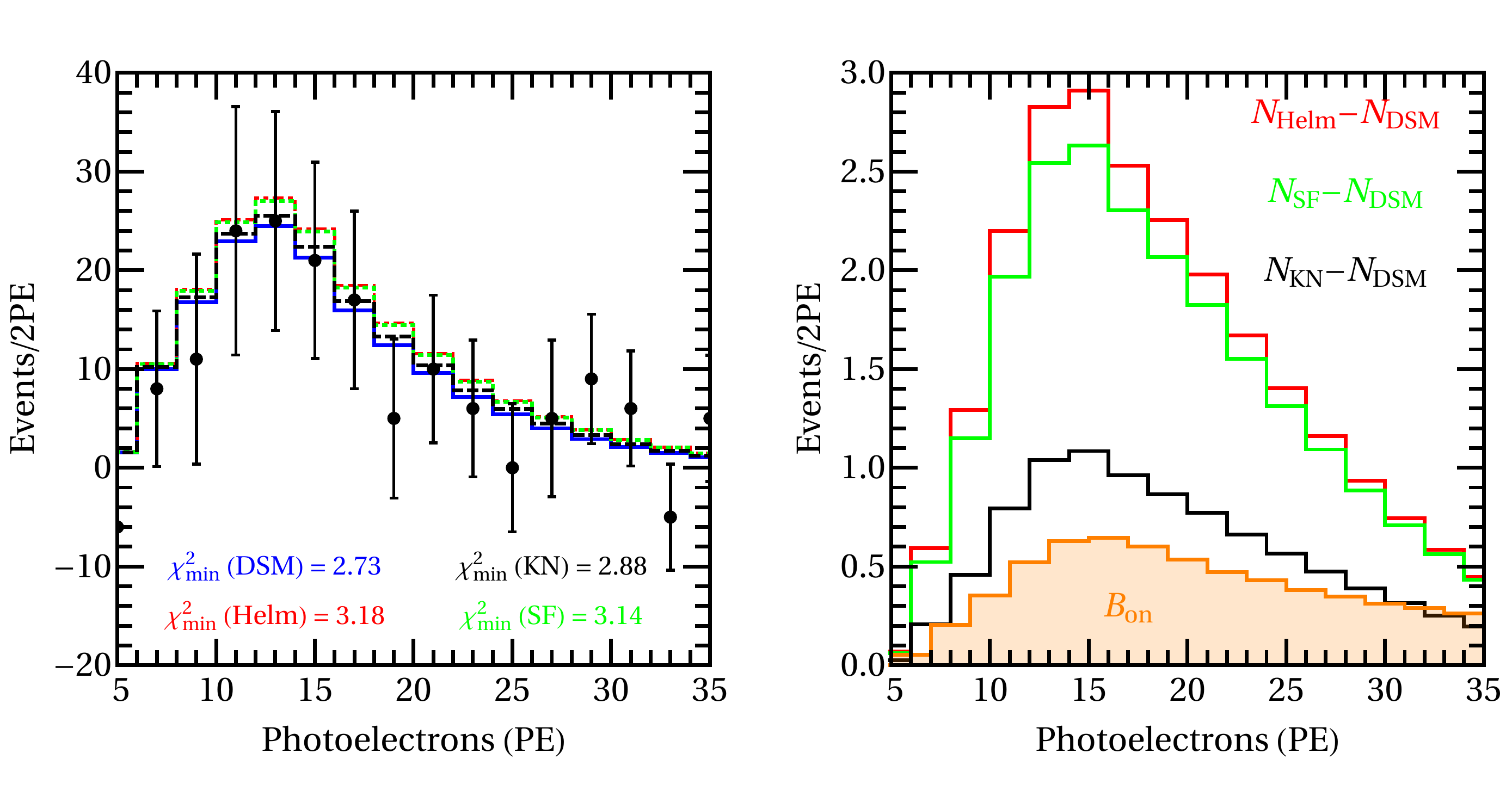}
\caption{Number of events per 2 photoelectrons at the COHERENT experiment. Left:
comparison of the corresponding results calculated with DSM and conventional
Helm, SF, KN form factors and the experimental data. Right: difference in events
between DSM and phenomenological form factor calculations and Beam-on prompt
neutron background events as a function of observed photoelectrons. For details
see the text.}
\label{fig:SNS-events}
\end{figure*}

The main results of the present work come out of a statistical analysis
of the COHERENT data through the $\chi ^{2}$ function taken from
Ref.~\cite{Akimov:2017ade}
%
\begin{equation}
\begin{aligned}
\chi ^{2}
& (\mathcal{S}) = \\ & \underset{\xi , \zeta }{\mathrm{min}}
\Bigg [ \sum _{i=4}^{15}\frac{\left (N^{i}_{\mathrm{meas}} - N^{i}_{
\mathrm{theor}}(\mathcal{S}) [1+\xi ] - B_{0n}^{i} [1+\zeta ] \right )
^{2}}{(\sigma ^{i}_{\mathrm{stat}})^{2}} \\
& \quad + \left (\frac{
\xi }{\sigma _{\xi }} \right )^{2} + \left (\frac{\zeta }{\sigma _{
\zeta }} \right )^{2} \Bigg ] \, ,
\end{aligned}
\label{eq:chi}
\end{equation}
where $\xi $ and $\zeta $ are the systematic parameters to account for
the uncertainties on the signal and background rates respectively, with
fractional uncertainties $\sigma _{\xi }= 0.28$ and $\sigma _{\zeta }=
0.25$. The quantities $B_{0n}^{i}$ and $\sigma ^{i}_{\mathrm{stat}}$
denote the $i$-th bin of the beam-on prompt neutron background events
and the statistical uncertainty respectively (see
Ref.~\cite{Akimov:2017ade} for details). Here, $B_{0n}^{i}$ is
evaluated by weighting the available experimental values from the
COHERENT data release~\cite{Akimov:2018vzs} with the total energy
delivered during the first run e.g. 7.47594 GWhr and the detector
efficiency (see also Ref.~\cite{AristizabalSierra:2018eqm}). In
Eq.~(\ref{eq:chi}), $\mathcal{S}$ represents the set of parameters for
which our theoretical calculation on $N_{\mathrm{theor}}(\mathcal{S})$
is evaluated. By minimizing over the nuisance parameters, we fit the
COHERENT data and calculate $\Delta \chi ^{2}(\mathcal{S})= \chi ^{2}(
\mathcal{S}) - \chi ^{2}_{\mathrm{min}}(\mathcal{S})$ which allows us to
probe the nuclear parameters in question. Finally, in our calculations
we restrict ourselves in the region $6 \leq n_{\mathrm{PE}} \leq 30$
corresponding to 12 energy bins in the range $4 \leq \mathrm{bin}
\leq 15$.

%

\begin{table*}[ht!]
\centering
\begin{tabular}{lcccc}
\hline
COHERENT & $\mathcal{F}^{\prime }/\mathcal{F}$ & Stat. uncertainty & Syst. uncertainty & $\langle R^{2}_{n}\rangle ^{1/2}$ (fm) \\
\hline
phase I & 1 & current~\cite{Akimov:2017ade} & current~\cite{Akimov:2017ade} & $5.64^{+ 0.99}_{-1.23}$\\ [3pt]
scenario I & 10 & $\sigma _{\text{stat}}=0.2 $ & $\sigma _{\text{sys}}=0.14$ & $5.23^{+ 0.42}_{-0.50}$ \\[3pt]
scenario II & 100 & $\sigma _{\text{stat}}=0.1 $ & $\sigma _{\text{sys}}=0.07$ & $5.23^{+ 0.22}_{-0.22}$ \\
\hline
\end{tabular}
\caption{Current and future experimental setups considered in the present study
and fitted neutron rms radii.}
\label{tab:SNS-upgrades}
\end{table*}

The aforementioned discrepancy between the DSM and the conventional
Helm, SF and KN form factors motivates us to conduct a more systematic
study of the relevant nuclear physics parameters.
 Fig.~\ref{fig:SNS-events} illustrates the estimated number of events
within DSM, and compares the recent COHERENT data with the calculations
considering the phenomenological form factors. From the left panel of
this figure it can be seen that an improved agreement with the
experimental data is found in the context of the employed realistic DSM
calculations. Indeed, our present DSM calculations result to a better
fit of the experimental data with $\chi ^{2}_{\text{min}}(\text{DSM})=2.73$
compared to $\chi ^{2}_{\text{min}}(\text{Helm})=3.18$,
$\chi ^{2}_{\text{min}}(\text{SF})=3.14$ and $\chi ^{2}_{\text{min}}(\text{KN})=2.88$
evaluated in the framework of a Helm, SF and KN form factor
approximations.

As demonstrated in Ref.~\cite{AristizabalSierra:2018eqm} the
resulted fit allows to accommodate  new physics and therefore advanced
nuclear physics models such as the DSM are essential for beyond the SM
searches too. Despite the fact that this difference lies well within the
present experimental error, we stress that future precise measurements
expected during the next phases of COHERENT~\cite{Akimov:2018ghi}
or from the upcoming CE$\nu $NS reactor
experiments~\cite{Wong:2010zzc,Aguilar-Arevalo:2016qen,Agnolet:2016zir,Belov:2015ufh,conus,Billard:2016giu,Strauss:2017cuu}
motivate the adoption of realistic nuclear structure methods especially
for the accurate characterization of the nuclear target responses. For
illustration purposes, the right panel of  Fig.~\ref{fig:SNS-events}
depicts the difference in events between the DSM and each of the
conventional form factor calculations e.g. $N_{\text{Helm}}-N_{
\text{DSM}}$, $N_{\text{SF}}-N_{\text{DSM}}$ and $N_{\text{KN}}-N_{
\text{DSM}}$ compared to the beam-on prompt neutron background events
$B_{0n}$ as functions of the detected photoelectrons. For completeness,
we note that the differences in events between the Helm and SF form
factor calculations (not shown here) are lower than the $B_{0n}$ level.

We now focus on the current potential of the COHERENT experiment to
probe important ingredients of the nuclear form factors in question. The
next stages of COHERENT experiment include future upgrades with
Germanium, LAr and NaI[Tl] detectors with mass up to
ton-scale~\cite{Akimov:2018vzs} that will not be considered in our
study (we are mainly interested in the study of Cs and I isotopes). The
CsI detector subsystem will continue to take data and the COHERENT
Collaboration aims to reduce the statistical
uncertainties~\cite{Akimov:2018vzs}. We are therefore motivated
to explore the attainable future sensitivities by assuming two possible
upgrades, namely scenario I and II. The number of events is scaled up
in terms of the factor $\mathcal{F}^{\prime }$ that quantifies the
exposure time, the detector mass and the SNS beam power [see
Eq.~(\ref{eq:F})] while, following Ref.~\cite{Cadeddu:2017etk}, we
choose an improved statistical/systematic uncertainty. Specifically, we
consider (i) a conservative future scenario I with $\mathcal{F}^{
\prime }/\mathcal{F}=10$ and half systematic uncertainty compared to
COHERENT first run, and (ii) an optimistic future scenario II with
$\mathcal{F}^{\prime }/\mathcal{F}=100$ and a systematic uncertainty
that is 25\% of the first phase of COHERENT. For the statistical
uncertainty in each case and more details see
 Table~\ref{tab:SNS-upgrades}. Finally, in order to cover future
scenarios, our calculations rely on the following $\chi ^{2}$ function
%
\begin{equation}
\begin{aligned}
\chi ^{2}(\mathcal{S}) = \underset{\xi }{\mathrm{min}} \Bigg [ \frac{
\left (N_{\text{DSM}} - N_{\text{theor}}(\mathcal{S}) [1+\xi ]\right )
^{2}}{N_{\text{DSM}}(1 + \sigma _{\text{stat}})} + \left ( \frac{
\xi }{\sigma _{\text{sys}}}\right )^{2} \Bigg ] \, ,
\end{aligned}
\end{equation}
where in this case $N_{\text{DSM}}$ denotes the number of events
predicted within the context of the DSM.

%
\begin{figure}
\includegraphics[width=0.9\linewidth]{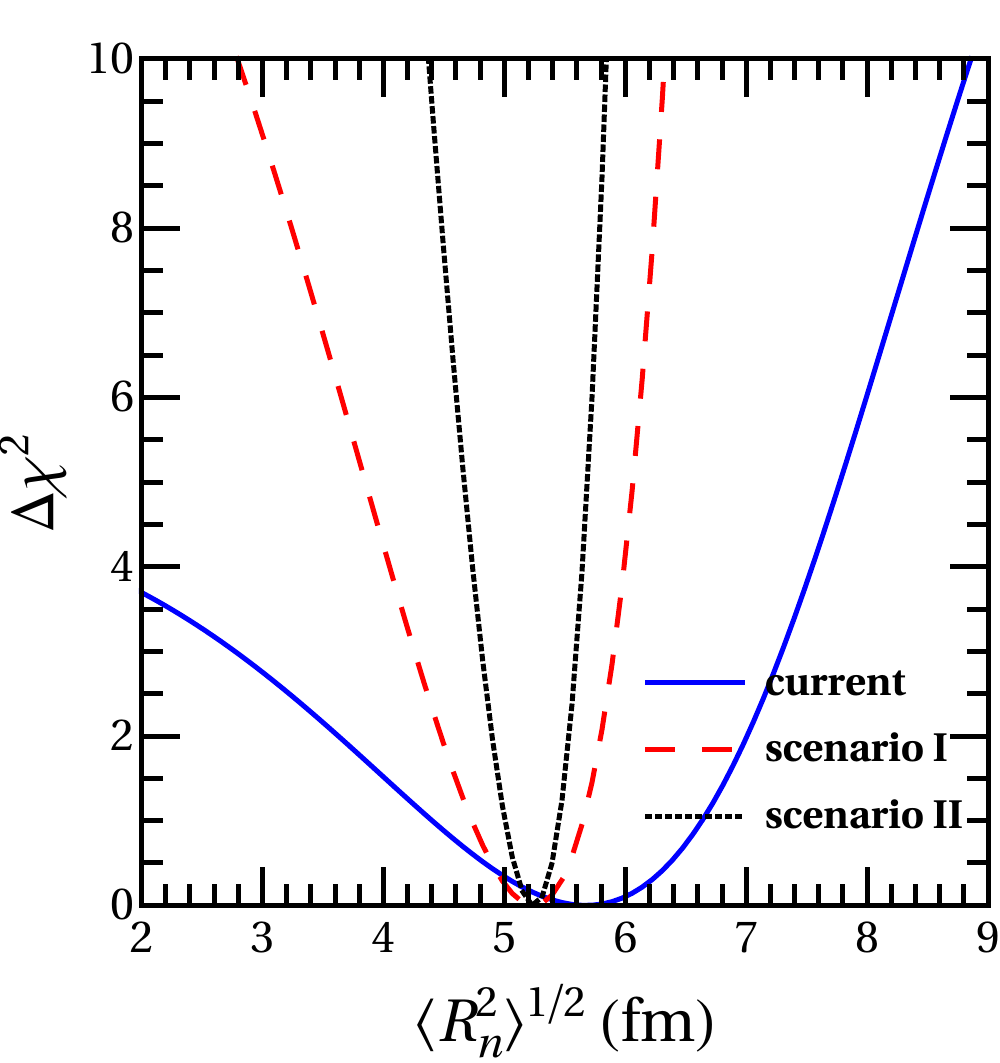}
\caption{$\Delta \chi ^{2}$ profile of the neutron rms radius $\langle
R_{n}^{2}\rangle ^{1/2}$. The results are presented for different experimental
setups.}
\label{fig:h.o.}
\end{figure}

In Ref.~\cite{Cadeddu:2017etk} it is shown that the recent CE$
\nu $NS data offer a unique pathway to probe the neutron rms radius. We
perform a sensitivity analysis based on the corresponding $\chi ^{2}
\left (\langle R_{n}^{2} \rangle ^{1/2}\right )$ function and our present
results are depicted in  Fig.~\ref{fig:h.o.}. For the current phase we
find the best fit value $\langle R_{n}^{2} \rangle ^{1/2} = 5.64^{+ 0.99}
_{-1.23}$ fm in good agreement with
Refs.~\cite{Cadeddu:2017etk,Huang:2019ene} (see
 Table~\ref{tab:SNS-upgrades}), while the results do not depend
significantly on the form factor used. Then, exploring the capability
of a future COHERENT experiment with upgrades according to scenarios I
and II we find the respective values $5.23^{+ 0.42}_{-0.50}$ fm in
scenario I and $5.23^{+ 0.22}_{-0.22}$ fm in scenario II. From the
latter we extract the conclusion that future COHERENT data alone (see
Ref.~\cite{Akimov:2018ghi} for details), will offer a better
determination of $\langle R_{n}^{2} \rangle ^{1/2} $ compared to the
current best limit reported in Ref.~\cite{Cadeddu:2018izq} that
was obtained through a combined analysis of the available CE$\nu $NS and
APV in Cs data. It is worth mentioning that such results remain
essentially unaltered regardless of the form factor used (see also
Ref.~\cite{Cadeddu:2017etk}). We finally stress that the present
work involves weak charge nuclear radii obtained from the coherent data.
We note however, that a more accurate comparison with the point nucleon
radii involves  the ``weak charge skin''~\cite{Horowitz:2012tj}.

\begin{figure*}
\centering
\includegraphics[width=0.9\textwidth]{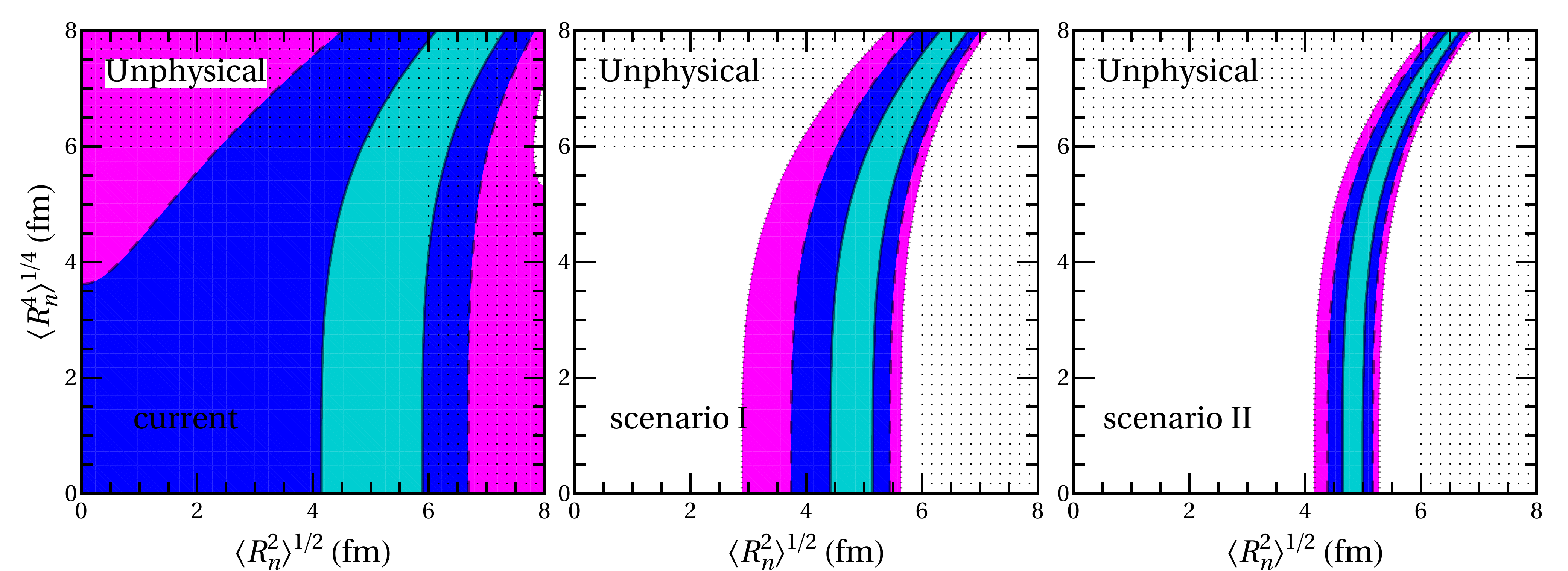}
\caption{Allowed regions in the $\langle R_{n}^{2} \rangle ^{1/2} $--$ \langle
R_{n}^{4} \rangle ^{1/4}$ parameter space from the COHERENT data for different
detector specifications (see the text). The contours correspond to $1\sigma $
(turquoise), 90\protect\% C.L. (blue) and 99\protect\% C.L. (magenta).}
\label{fig:RS}
\end{figure*}

We now consider the model independent expansion of the form factor given
in Eq.~(\ref{eq:form-factor-expansion}). In what follows, we will
consider only the neutron form factor which dominates the CE$\nu $NS
process. For simplicity we take into account only the two first (even)
moments and perform a combined sensitivity analysis of the current and
future COHERENT data on the basis of the $\chi ^{2}\left (\langle R
_{n}^{2} \rangle ,\langle R_{n}^{4} \rangle \right )$ function. In this
calculation we restrict ourselves in the physical region [0,6] fm that
is determined from the upper limit on $ R_{n}(^{208} \text{Pb})=5.75
\pm 0.18$ fm from the PREM experiment~\cite{Horowitz:2012tj} (see
also Ref.~\cite{AristizabalSierra:2019zmy}). The corresponding
bounds are shown in  Fig.~\ref{fig:RS} at $1\sigma $, 90\% and 99\% C.L.
The constraints are not yet competitive to current experimental
results~\cite{Angeli:2013epw}, while there are prospects of
significant improvement in future measurements according to scenarios
I and II. It can also be seen that the 4-th moment, $\langle R_{n}
^{4}\rangle $, under the assumptions of the present study is not well
constrained. We however emphasize that largely improved constraints are
possible at multi-ton scale CE$\nu $NS
detectors~\cite{Patton:2012jr}.

\begin{figure*}[t]
\centering
\includegraphics[width=0.9\textwidth]{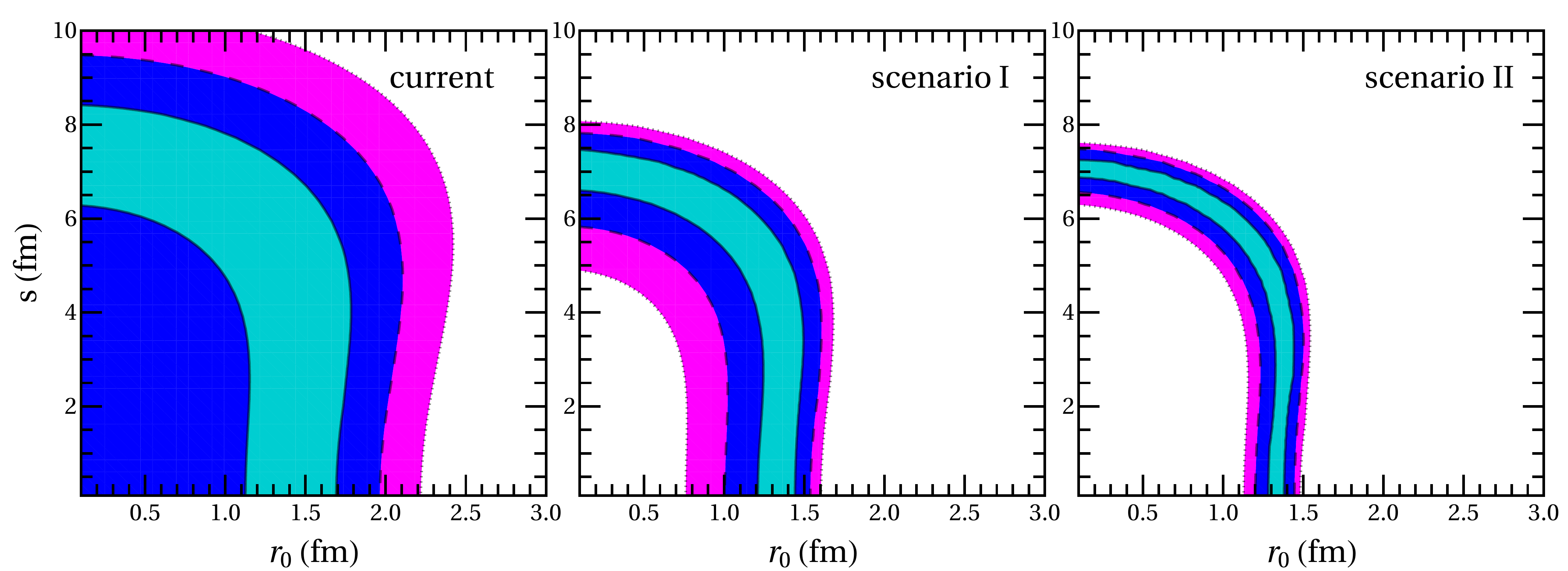}
\includegraphics[width=0.9\textwidth]{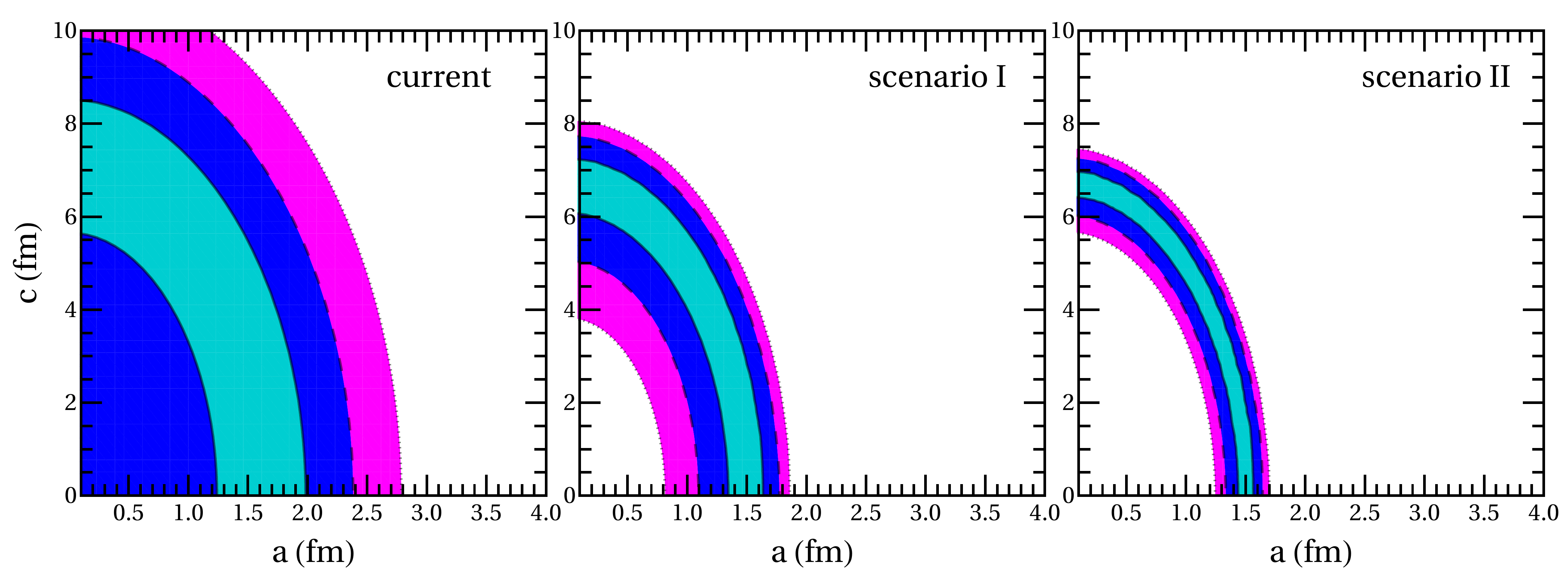}
\includegraphics[width=0.9\textwidth]{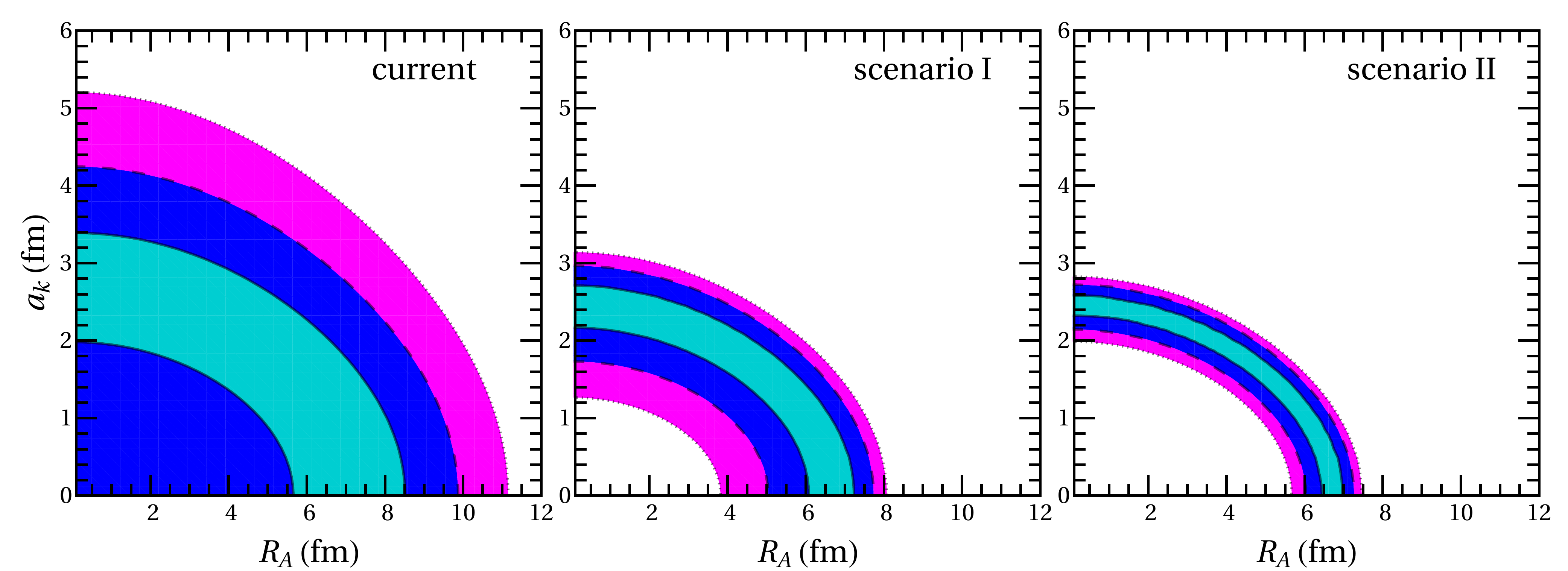}
\caption{Allowed regions in the $r_{0}$--$s$ (top), in the $a$--$c$ (middle) and
in the $R_{A}$--$a_{k}$ (bottom) parameter space from the COHERENT data,
corresponding to Helm, SF and KN form factors respectively. Different detector
specifications have been considered (see the text). The results are presented at
$1\sigma $, 90\% C.L. and 99\% C.L. (for the color coding see
Fig.~\ref{fig:RS}).}
\label{fig:form_factors_combined}
\end{figure*}

It is now worthwhile to explore the possibility of extracting
simultaneous constraints on the parameters characterizing the Helm, SF
and KN form factors, from CE$\nu $NS data. In our aim to explore the
Helm form factor given in Eq.~(\ref{eq:helm}), we consider the
parameterization $F_{\text{Helm}} \left (Q^{2}, r_{0}, s \right )$ with
diffraction radius $R_{0}= r_{0} A^{1/3}$ and we perform a 2-parameter
fit based on the $\chi ^{2} \left (r_{0},s \right )$ function. The
allowed regions in the $(r_{0},s)$ plane are illustrated in the upper
panel of  Fig.~\ref{fig:form_factors_combined} at $1\sigma $, 90\% and
99\% C.L., under the assumptions of the current (phase I) and the
scenarios I and II. Although it becomes evident that future measurements
will drastically improve the current constraints, it can be seen that
CE$\nu $NS data are not sensitive to the surface thickness, $s$. This
conclusion is in agreement with a recent study of
Ref.~\cite{AristizabalSierra:2019zmy}, while the prospect of
probing $r_{0}$ is significant.

For the case of the SF form factor, we explore the allowed region in the
$(a,c)$ parameter space. By marginalizing the relevant $\chi ^{2}(a,c)$
function, we present the contours of the half-density radius $c$ with
the surface diffuseness $a$ at $1\sigma $, 90\% and 99\% C.L in the
middle panel of  Fig.~\ref{fig:form_factors_combined}. The present
results imply that in a future COHERENT experiment, the prospects of
improvement with respect to the current constraints are rather promising
and can be competitive with existing
analyses~\cite{Horowitz:2012tj,Piekarewicz:2016vbn} on
$^{208}$Pb from PREX data~\cite{Abrahamyan:2012gp}.

In a similar way, we explore the attainable constraints on the
$(R_{A},a_{k})$ parameters entering the KN form factor. In this case,
the $1\sigma $, 90\% and 99\% C.L allowed regions are depicted in the
lower panel of  Fig.~\ref{fig:form_factors_combined}. Likewise, there is
a large potential of improvement from future CE$\nu $NS measurements
during the next phases of the COHERENT program. Finally, we perform a
sensitivity fit based on the following parametrization of the effective
nuclear radius~\cite{Orrigo:2016wgu}
%
\begin{equation}
R= r_{0} A^{1/3} + r_{1} \, .
\end{equation}
Marginalizing over $r_{1}$, we find the best fit values
%
\begin{equation}
\begin{aligned}
r_{0}=
& 1.28^{+0.58}_{-0.58}, \quad \text{current}\, ,\\
r_{0}=& 1.23^{+0.37}
_{-0.27}, \quad \text{scenario I}\, ,\\
r_{0}=& 1.23^{+0.31}_{-0.20},
\quad \text{scenario II} \, ,
\end{aligned}
\end{equation}
being consistent with Eq.~(\ref{eq:SF-vals}) and
Ref.~\cite{Lewin:1995rx}.


\section{Conclusions}%
\label{sec:conclusion}

The present work, relying on improved nuclear structure calculations
employing DSM that starts with the same shell model inputs, gives a
better interpretation of the current and future COHERENT data in which
a large portion of the theoretical uncertainty originates from the
calculation of the neutron nuclear form factors. We devoted a thorough
analysis on the available CE$\nu $NS data and extracted constraints to
the nuclear parameters characterizing the Helm, symmetrized Fermi and
Klein-Nystrand form factor distributions. We also investigated the near-
and long-term future sensitivities, within the context of two possible
scenarios, and concluded that there is a large potential of improvement.
We have checked that the constraints on the nuclear rms radius do not
essentially depend on the form factor choice that is used to analyze the
data. Moreover, we have shown that future COHERENT measurements alone
will reach a better sensitivity on the neutron rms radius compared to
the best current limits that were recently extracted from a combined
analysis of the available data from CE$\nu $NS and APV data. Finally we
have presented simultaneous constraints on the parameters characterizing
the phenomenological form factors as well as for the first two moments
of the neutron form factor (the sensitivity of the form factor on
pairing and deformation will be studied in detail in a separate work).
Reducing the latter uncertainty, possible deviations from the SM
expectations may be extracted with high significance.

\acknowledgements 
\noindent
DKP has been supported by the Spanish grants SEV-2014-0398 and FPA2017-85216-P (AEI/FEDER, UE), PROMETEO/2018/165 (Generalitat Valenciana) and the Spanish Red Consolider MultiDark FPA2017-90566-REDC. RS is thankful to SERB of Department of Science and Technology (Government of India) for financial support. DKP acknowledges stimulating discussions with K. Patton, C. Giunti and M. T\'ortola.

\bibliographystyle{utphys}
\bibliography{final_plb.bbl}

\providecommand{\href}[2]{#2}\begingroup\raggedright\begin{thebibliography}{10}

\bibitem{Akimov:2017ade}
{\bfseries COHERENT} Collaboration, D.~Akimov {\em et~al.}, ``{Observation of
  Coherent Elastic Neutrino-Nucleus Scattering},''
  \href{http://dx.doi.org/10.1126/science.aao0990}{{\em Science} {\bfseries
  357} (2017) 1123--1126}, \href{http://arxiv.org/abs/1708.01294}{{\ttfamily
  arXiv:1708.01294 [nucl-ex]}}.

\bibitem{Akimov:2018vzs}
{\bfseries COHERENT} Collaboration, D.~Akimov {\em et~al.}, ``{COHERENT
  Collaboration data release from the first observation of coherent elastic
  neutrino-nucleus scattering},''
  \href{http://arxiv.org/abs/1804.09459}{{\ttfamily arXiv:1804.09459
  [nucl-ex]}}.

\bibitem{Ejiri:2019ezh}
H.~Ejiri, J.~Suhonen, and K.~Zuber, ``{Neutrino-nuclear responses for
  astro-neutrinos, single beta decays and double beta decays},''
  \href{http://dx.doi.org/10.1016/j.physrep.2018.12.001}{{\em Phys. Rept.}
  {\bfseries 797} (2019) 1--102}.

\bibitem{Kosmas:2017tsq}
D.~Papoulias and T.~Kosmas, ``{COHERENT constraints to conventional and exotic
  neutrino physics},'' \href{http://dx.doi.org/10.1103/PhysRevD.97.033003}{{\em
  Phys.Rev.} {\bfseries D97} (2018) 033003},
  \href{http://arxiv.org/abs/1711.09773}{{\ttfamily arXiv:1711.09773
  [hep-ph]}}.

\bibitem{Barranco:2005yy}
J.~Barranco, O.~Miranda, and T.~Rashba, ``{Probing new physics with coherent
  neutrino scattering off nuclei},''
  \href{http://dx.doi.org/10.1088/1126-6708/2005/12/021}{{\em JHEP} {\bfseries
  0512} (2005) 021}.

\bibitem{Scholberg:2005qs}
K.~Scholberg, ``{Prospects for measuring coherent neutrino-nucleus elastic
  scattering at a stopped-pion neutrino source},''
  \href{http://dx.doi.org/10.1103/PhysRevD.73.033005}{{\em Phys.Rev.}
  {\bfseries D73} (2006) 033005}.

\bibitem{Liao:2017uzy}
J.~Liao and D.~Marfatia, ``{COHERENT constraints on nonstandard neutrino
  interactions},'' \href{http://dx.doi.org/10.1016/j.physletb.2017.10.046}{{\em
  Phys.Lett.} {\bfseries B775} (2017) 54--57},
  \href{http://arxiv.org/abs/1708.04255}{{\ttfamily arXiv:1708.04255
  [hep-ph]}}.

\bibitem{Dent:2017mpr}
J.~B. Dent, B.~Dutta, S.~Liao, J.~L. Newstead, L.~E. Strigari, and J.~W.
  Walker, ``{Accelerator and reactor complementarity in coherent
  neutrino-nucleus scattering},''
  \href{http://dx.doi.org/10.1103/PhysRevD.97.035009}{{\em Phys.Rev.}
  {\bfseries D97} (2018) 035009},
  \href{http://arxiv.org/abs/1711.03521}{{\ttfamily arXiv:1711.03521
  [hep-ph]}}.

\bibitem{AristizabalSierra:2017joc}
D.~Aristizabal~Sierra, N.~Rojas, and M.~Tytgat, ``{Neutrino non-standard
  interactions and dark matter searches with multi-ton scale detectors},''
  \href{http://dx.doi.org/10.1007/JHEP03(2018)197}{{\em JHEP} {\bfseries 1803}
  (2018) 197}, \href{http://arxiv.org/abs/1712.09667}{{\ttfamily
  arXiv:1712.09667 [hep-ph]}}.

\bibitem{Denton:2018xmq}
P.~B. Denton, Y.~Farzan, and I.~M. Shoemaker, ``{Testing large non-standard
  neutrino interactions with arbitrary mediator mass after COHERENT data},''
  \href{http://dx.doi.org/10.1007/JHEP07(2018)037}{{\em JHEP} {\bfseries 1807}
  (2018) 037}, \href{http://arxiv.org/abs/1804.03660}{{\ttfamily
  arXiv:1804.03660 [hep-ph]}}.

\bibitem{Kosmas:2015sqa}
T.~Kosmas, O.~Miranda, D.~Papoulias, M.~Tortola, and J.~Valle, ``{Probing
  neutrino magnetic moments at the Spallation Neutron Source facility},''
  \href{http://dx.doi.org/10.1103/PhysRevD.92.013011}{{\em Phys.Rev.}
  {\bfseries D92} (2015) 013011},
  \href{http://arxiv.org/abs/1505.03202}{{\ttfamily arXiv:1505.03202
  [hep-ph]}}.

\bibitem{Billard:2018jnl}
J.~Billard, J.~Johnston, and B.~J. Kavanagh, ``{Prospects for exploring New
  Physics in Coherent Elastic Neutrino-Nucleus Scattering},''
  \href{http://dx.doi.org/10.1088/1475-7516/2018/11/016}{{\em JCAP} {\bfseries
  1811} (2018) 016}, \href{http://arxiv.org/abs/1805.01798}{{\ttfamily
  arXiv:1805.01798 [hep-ph]}}.

\bibitem{Cadeddu:2018dux}
M.~Cadeddu, C.~Giunti, K.~Kouzakov, Y.~Li, A.~Studenikin, and Y.~Zhang,
  ``{Neutrino Charge Radii from COHERENT Elastic Neutrino-Nucleus
  Scattering},'' \href{http://dx.doi.org/10.1103/PhysRevD.98.113010}{{\em
  Phys.Rev.} {\bfseries D98} (2018) 113010},
  \href{http://arxiv.org/abs/1810.05606}{{\ttfamily arXiv:1810.05606
  [hep-ph]}}.

\bibitem{Kosmas:2017zbh}
T.~Kosmas, D.~Papoulias, M.~Tortola, and J.~Valle, ``{Probing light sterile
  neutrino signatures at reactor and Spallation Neutron Source neutrino
  experiments},'' \href{http://dx.doi.org/10.1103/PhysRevD.96.063013}{{\em
  Phys.Rev.} {\bfseries D96} (2017) 063013},
  \href{http://arxiv.org/abs/1703.00054}{{\ttfamily arXiv:1703.00054
  [hep-ph]}}.

\bibitem{Canas:2017umu}
B.~Ca{\~n}as, E.~Garc{\'e}s, O.~Miranda, and A.~Parada, ``{The reactor
  antineutrino anomaly and low energy threshold neutrino experiments},''
  \href{http://dx.doi.org/10.1016/j.physletb.2017.11.074}{{\em Phys.Lett.}
  {\bfseries B776} (2018) 451--456},
  \href{http://arxiv.org/abs/1708.09518}{{\ttfamily arXiv:1708.09518
  [hep-ph]}}.

\bibitem{Blanco:2019vyp}
C.~Blanco, D.~Hooper, and P.~Machado, ``{Constraining Sterile Neutrino
  Interpretations of the LSND and MiniBooNE Anomalies with Coherent Neutrino
  Scattering Experiments},'' \href{http://arxiv.org/abs/1901.08094}{{\ttfamily
  arXiv:1901.08094 [hep-ph]}}.

\bibitem{Dent:2016wcr}
J.~B. Dent, B.~Dutta, S.~Liao, J.~L. Newstead, L.~E. Strigari, and J.~W.
  Walker, ``{Probing light mediators at ultralow threshold energies with
  coherent elastic neutrino-nucleus scattering},''
  \href{http://dx.doi.org/10.1103/PhysRevD.96.095007}{{\em Phys.Rev.}
  {\bfseries D96} (2017) 095007},
  \href{http://arxiv.org/abs/1612.06350}{{\ttfamily arXiv:1612.06350
  [hep-ph]}}.

\bibitem{Farzan:2018gtr}
Y.~Farzan, M.~Lindner, W.~Rodejohann, and X.-J. Xu, ``{Probing neutrino
  coupling to a light scalar with coherent neutrino scattering},''
  \href{http://dx.doi.org/10.1007/JHEP05(2018)066}{{\em JHEP} {\bfseries 1805}
  (2018) 066}, \href{http://arxiv.org/abs/1802.05171}{{\ttfamily
  arXiv:1802.05171 [hep-ph]}}.

\bibitem{Abdullah:2018ykz}
M.~Abdullah, J.~B. Dent, B.~Dutta, G.~L. Kane, S.~Liao, and L.~E. Strigari,
  ``{Coherent elastic neutrino nucleus scattering as a probe of a $Z^\prime$
  through kinetic and mass mixing effects},''
  \href{http://dx.doi.org/10.1103/PhysRevD.98.015005}{{\em Phys.Rev.}
  {\bfseries D98} (2018) 015005},
  \href{http://arxiv.org/abs/1803.01224}{{\ttfamily arXiv:1803.01224
  [hep-ph]}}.

\bibitem{Brdar:2018qqj}
V.~Brdar, W.~Rodejohann, and X.-J. Xu, ``{Producing a new Fermion in Coherent
  Elastic Neutrino-Nucleus Scattering: from Neutrino Mass to Dark Matter},''
  \href{http://dx.doi.org/10.1007/JHEP12(2018)024}{{\em JHEP} {\bfseries 1812}
  (2018) 024}, \href{http://arxiv.org/abs/1810.03626}{{\ttfamily
  arXiv:1810.03626 [hep-ph]}}.

\bibitem{Miranda:2019wdy}
O.~Miranda, D.~Papoulias, M.~T{\'o}rtola, and J.~Valle, ``{Probing neutrino
  transition magnetic moments with coherent elastic neutrino-nucleus
  scattering},'' \href{http://dx.doi.org/10.1007/JHEP07(2019)103}{{\em JHEP}
  {\bfseries 1907} (2019) 103},
  \href{http://arxiv.org/abs/1905.03750}{{\ttfamily arXiv:1905.03750
  [hep-ph]}}.

\bibitem{AristizabalSierra:2019ufd}
D.~Aristizabal~Sierra, V.~De~Romeri, and N.~Rojas, ``{CP violating effects in
  coherent elastic neutrino-nucleus scattering processes},''
  \href{http://dx.doi.org/10.1007/JHEP09(2019)069}{{\em JHEP} {\bfseries 09}
  (2019) 069},
\href{http://arxiv.org/abs/1906.01156}{{\ttfamily arXiv:1906.01156 [hep-ph]}}.

\bibitem{Ge:2017mcq}
S.-F. Ge and I.~M. Shoemaker, ``{Constraining Photon Portal Dark Matter with
  Texono and Coherent Data},''
  \href{http://dx.doi.org/10.1007/JHEP11(2018)066}{{\em JHEP} {\bfseries 1811}
  (2018) 066}, \href{http://arxiv.org/abs/1710.10889}{{\ttfamily
  arXiv:1710.10889 [hep-ph]}}.

\bibitem{Coloma:2017ncl}
P.~Coloma, M.~Gonzalez-Garcia, M.~Maltoni, and T.~Schwetz, ``{COHERENT
  Enlightenment of the Neutrino Dark Side},''
  \href{http://dx.doi.org/10.1103/PhysRevD.96.115007}{{\em Phys.Rev.}
  {\bfseries D96} (2017) 115007},
  \href{http://arxiv.org/abs/1708.02899}{{\ttfamily arXiv:1708.02899
  [hep-ph]}}.

\bibitem{Gonzalez-Garcia:2018dep}
M.~Gonzalez-Garcia, M.~Maltoni, Y.~F. Perez-Gonzalez, and
  R.~Zukanovich~Funchal, ``{Neutrino Discovery Limit of Dark Matter Direct
  Detection Experiments in the Presence of Non-Standard Interactions},''
  \href{http://dx.doi.org/10.1007/JHEP07(2018)019}{{\em JHEP} {\bfseries 1807}
  (2018) 019}, \href{http://arxiv.org/abs/1803.03650}{{\ttfamily
  arXiv:1803.03650 [hep-ph]}}.

\bibitem{Ng:2017aur}
K.~C.~Y. Ng, J.~F. Beacom, A.~H.~G. Peter, and C.~Rott, ``{Solar Atmospheric
  Neutrinos: A New Neutrino Floor for Dark Matter Searches},''
  \href{http://dx.doi.org/10.1103/PhysRevD.96.103006}{{\em Phys.Rev.}
  {\bfseries D96} (2017) 103006},
  \href{http://arxiv.org/abs/1703.10280}{{\ttfamily arXiv:1703.10280
  [astro-ph.HE]}}.

\bibitem{Papoulias:2018uzy}
D.~Papoulias, R.~Sahu, T.~Kosmas, V.~Kota, and B.~Nayak, ``{Novel
  neutrino-floor and dark matter searches with deformed shell model
  calculations},'' \href{http://dx.doi.org/10.1155/2018/6031362}{{\em Adv.High
  Energy Phys.} {\bfseries 2018} (2018) 6031362},
  \href{http://arxiv.org/abs/1804.11319}{{\ttfamily arXiv:1804.11319
  [hep-ph]}}.

\bibitem{Boehm:2018sux}
C.~B{\oe}hm, D.~Cerde{\~n}o, P.~N. Machado, A.~Olivares-Del~Campo, and E.~Reid,
  ``{How high is the neutrino floor?},''
  \href{http://dx.doi.org/10.1088/1475-7516/2019/01/043}{{\em JCAP} {\bfseries
  1901} (2019) 043}, \href{http://arxiv.org/abs/1809.06385}{{\ttfamily
  arXiv:1809.06385 [hep-ph]}}.

\bibitem{Dutta:2019oaj}
B.~Dutta and L.~E. Strigari, ``{Neutrino physics with dark matter detectors},''
\href{http://arxiv.org/abs/1901.08876}{{\ttfamily arXiv:1901.08876 [hep-ph]}}.

\bibitem{Bednyakov:2018mjd}
V.~A. Bednyakov and D.~V. Naumov, ``{Coherency and incoherency in
  neutrino-nucleus elastic and inelastic scattering},''
  \href{http://dx.doi.org/10.1103/PhysRevD.98.053004}{{\em Phys.Rev.}
  {\bfseries D98} (2018) 053004},
  \href{http://arxiv.org/abs/1806.08768}{{\ttfamily arXiv:1806.08768
  [hep-ph]}}.

\bibitem{Lindner:2016wff}
M.~Lindner, W.~Rodejohann, and X.-J. Xu, ``{Coherent Neutrino-Nucleus
  Scattering and new Neutrino Interactions},''
  \href{http://dx.doi.org/10.1007/JHEP03(2017)097}{{\em JHEP} {\bfseries 1703}
  (2017) 097}, \href{http://arxiv.org/abs/1612.04150}{{\ttfamily
  arXiv:1612.04150 [hep-ph]}}.

\bibitem{AristizabalSierra:2018eqm}
D.~Aristizabal~Sierra, V.~De~Romeri, and N.~Rojas, ``{COHERENT analysis of
  neutrino generalized interactions},''
  \href{http://dx.doi.org/10.1103/PhysRevD.98.075018}{{\em Phys.Rev.}
  {\bfseries D98} (2018) 075018},
  \href{http://arxiv.org/abs/1806.07424}{{\ttfamily arXiv:1806.07424
  [hep-ph]}}.

\bibitem{Miranda:2019skf}
O.~G. Miranda, G.~Sanchez~Garcia, and O.~Sanders, ``{Coherent elastic
  neutrino-nucleus scattering as a precision test for the Standard Model and
  beyond: the COHERENT proposal case},''
  \href{http://dx.doi.org/10.1155/2019/3902819}{{\em Adv. High Energy Phys.}
  {\bfseries 2019} (2019) 3902819},
\href{http://arxiv.org/abs/1902.09036}{{\ttfamily arXiv:1902.09036 [hep-ph]}}.

\bibitem{Payne:2019wvy}
C.~Payne, S.~Bacca, G.~Hagen, W.~Jiang, and T.~Papenbrock, ``Coherent elastic
  neutrino-nucleus scattering on $^{40}$ar from first principles,''
  \href{http://arxiv.org/abs/1908.09739}{{\ttfamily arXiv:1908.09739
  [nucl-th]}}.

\bibitem{Cadeddu:2017etk}
M.~Cadeddu, C.~Giunti, Y.~Li, and Y.~Zhang, ``{Average CsI neutron density
  distribution from COHERENT data},''
  \href{http://dx.doi.org/10.1103/PhysRevLett.120.072501}{{\em Phys.Rev.Lett.}
  {\bfseries 120} (2018) 072501},
  \href{http://arxiv.org/abs/1710.02730}{{\ttfamily arXiv:1710.02730
  [hep-ph]}}.

\bibitem{Yang:2019pbx}
J.~Yang, J.~A. Hernandez, and J.~Piekarewicz, ``Electroweak probes of ground
  state densities,'' \href{http://dx.doi.org/10.1103/PhysRevC.100.054301}{{\em
  Phys.Rev.C} {\bfseries 100} no.~5, (2019) 054301},
  \href{http://arxiv.org/abs/1908.10939}{{\ttfamily arXiv:1908.10939
  [nucl-th]}}.

\bibitem{Smponias:2015nua}
T.~Smponias and O.~T. Kosmas, ``{High Energy Neutrino Emission from
  Astrophysical Jets in the Galaxy},''
\href{http://dx.doi.org/10.1155/2015/921757}{{\em Adv. High Energy Phys.}
  {\bfseries 2015} (2015) 921757}.

\bibitem{Fattoyev:2017jql}
F.~Fattoyev, J.~Piekarewicz, and C.~Horowitz, ``{Neutron Skins and Neutron
  Stars in the Multimessenger Era},''
  \href{http://dx.doi.org/10.1103/PhysRevLett.120.172702}{{\em Phys.Rev.Lett.}
  {\bfseries 120} (2018) 172702},
  \href{http://arxiv.org/abs/1711.06615}{{\ttfamily arXiv:1711.06615
  [nucl-th]}}.

\bibitem{ks-book}
V.~K.~B. Kota and R.~Sahu, {\em {Structure of Medium Mass Nuclei: Deformed
  Shell Model and Spin-Isospin Interacting Boson Model}}.
\newblock CRC Press, 2016.
\newblock
  \url{{https://www.amazon.com/Structure-Medium-Mass-Nuclei-Spin-Isospin-ebook/dp/B01MTZWTUT?SubscriptionId=0JYN1NVW651KCA56C102&tag=techkie-20&linkCode=xm2&camp=2025&creative=165953&creativeASIN=B01MTZWTUT}}.

\bibitem{Sahu:2013yna}
R.~Sahu, P.~Srivastava, and V.~Kota, ``{Deformed shell model results for
  neutrinoless positron double beta decay of nuclei in the A =
  60{\textendash}90 region},''
  \href{http://dx.doi.org/10.1088/0954-3899/40/9/095107}{{\em J.Phys.}
  {\bfseries G40} (2013) 095107}.

\bibitem{Sahu:2014nga}
R.~Sahu and V.~Kota, ``{Deformed shell model results for neutrinoless double
  beta decay of nuclei in A = 60 - 90 region},''
  \href{http://dx.doi.org/10.1142/S0218301315500226}{{\em Int.J.Mod.Phys.}
  {\bfseries E24} (2015) 1550022},
  \href{http://arxiv.org/abs/1409.4929}{{\ttfamily arXiv:1409.4929 [nucl-th]}}.

\bibitem{Kosmas:2003xr}
T.~S. Kosmas, A.~Faessler, and R.~Sahu, ``{Transition matrix elements for mu e
  conversion in ge-72 using the deformed Hartree-Fock method},''
  \href{http://dx.doi.org/10.1103/PhysRevC.68.054315}{{\em Phys.Rev.}
  {\bfseries C68} (2003) 054315}.

\bibitem{Sahu:2017czz}
R.~Sahu and V.~Kota, ``{Deformed shell model study of event rates for
  WIMP-$^{73}$Ge scattering},''
  \href{http://dx.doi.org/10.1142/S0217732317502108}{{\em Mod.Phys.Lett.}
  {\bfseries A32} (2017) 1750210},
  \href{http://arxiv.org/abs/1706.08112}{{\ttfamily arXiv:1706.08112
  [nucl-th]}}.

\bibitem{Papoulias:2015vxa}
D.~Papoulias and T.~Kosmas, ``{Standard and Nonstandard Neutrino-Nucleus
  Reactions Cross Sections and Event Rates to Neutrino Detection
  Experiments},'' \href{http://dx.doi.org/10.1155/2015/763648}{{\em Adv.High
  Energy Phys.} {\bfseries 2015} (2015) 763648},
  \href{http://arxiv.org/abs/1502.02928}{{\ttfamily arXiv:1502.02928
  [nucl-th]}}.

\bibitem{Pirinen:2018gsd}
P.~Pirinen, J.~Suhonen, and E.~Ydrefors, ``{Neutral-current neutrino-nucleus
  scattering off Xe isotopes},''
  \href{http://dx.doi.org/10.1155/2018/9163586}{{\em Adv.High Energy Phys.}
  {\bfseries 2018} (2018) 9163586},
  \href{http://arxiv.org/abs/1804.08995}{{\ttfamily arXiv:1804.08995
  [nucl-th]}}.

\bibitem{Kosmas:2015vsa}
T.~Kosmas, O.~Miranda, D.~Papoulias, M.~Tortola, and J.~Valle, ``{Sensitivities
  to neutrino electromagnetic properties at the TEXONO experiment},''
  \href{http://dx.doi.org/10.1016/j.physletb.2015.09.054}{{\em Phys.Lett.}
  {\bfseries B750} (2015) 459--465},
  \href{http://arxiv.org/abs/1506.08377}{{\ttfamily arXiv:1506.08377
  [hep-ph]}}.

\bibitem{Canas:2018rng}
B.~Ca{\~n}as, E.~Garc{\'e}s, O.~Miranda, and A.~Parada, ``{Future perspectives
  for a weak mixing angle measurement in coherent elastic neutrino nucleus
  scattering experiments},''
  \href{http://dx.doi.org/10.1016/j.physletb.2018.07.049}{{\em Phys.Lett.}
  {\bfseries B784} (2018) 159--162},
  \href{http://arxiv.org/abs/1806.01310}{{\ttfamily arXiv:1806.01310
  [hep-ph]}}.

\bibitem{Patton:2012jr}
K.~Patton, J.~Engel, G.~C. McLaughlin, and N.~Schunck, ``{Neutrino-nucleus
  coherent scattering as a probe of neutron density distributions},''
  \href{http://dx.doi.org/10.1103/PhysRevC.86.024612}{{\em Phys.Rev.}
  {\bfseries C86} (2012) 024612},
  \href{http://arxiv.org/abs/1207.0693}{{\ttfamily arXiv:1207.0693 [nucl-th]}}.

\bibitem{Ciuffoli:2018qem}
E.~Ciuffoli, J.~Evslin, Q.~Fu, and J.~Tang, ``{Extracting nuclear form factors
  with coherent neutrino scattering},''
  \href{http://dx.doi.org/10.1103/PhysRevD.97.113003}{{\em Phys.Rev.}
  {\bfseries D97} (2018) 113003},
  \href{http://arxiv.org/abs/1801.02166}{{\ttfamily arXiv:1801.02166
  [physics.ins-det]}}.

\bibitem{Akimov:2018ghi}
{\bfseries COHERENT} Collaboration, D.~Akimov {\em et~al.}, ``{COHERENT 2018 at
  the Spallation Neutron Source},''
  \href{http://arxiv.org/abs/1803.09183}{{\ttfamily arXiv:1803.09183
  [physics.ins-det]}}.

\bibitem{Wong:2010zzc}
H.~T. Wong, ``{Neutrino-nucleus coherent scattering and dark matter searches
  with sub-keV germanium detector},''
  \href{http://dx.doi.org/10.1016/j.nuclphysa.2010.05.040}{{\em Nucl.Phys.}
  {\bfseries A844} (2010) 229C--233C}.

\bibitem{Aguilar-Arevalo:2016qen}
{\bfseries CONNIE} Collaboration, A.~Aguilar-Arevalo {\em et~al.}, ``{Results
  of the Engineering Run of the Coherent Neutrino Nucleus Interaction
  Experiment (CONNIE)},''
  \href{http://dx.doi.org/10.1088/1748-0221/11/07/P07024}{{\em JINST}
  {\bfseries 11} (2016) P07024},
  \href{http://arxiv.org/abs/1604.01343}{{\ttfamily arXiv:1604.01343
  [physics.ins-det]}}.

\bibitem{Agnolet:2016zir}
{\bfseries MINER} Collaboration, G.~Agnolet {\em et~al.}, ``{Background Studies
  for the MINER Coherent Neutrino Scattering Reactor Experiment},''
  \href{http://dx.doi.org/10.1016/j.nima.2017.02.024}{{\em Nucl.Instrum.Meth.}
  {\bfseries A853} (2017) 53--60},
  \href{http://arxiv.org/abs/1609.02066}{{\ttfamily arXiv:1609.02066
  [physics.ins-det]}}.

\bibitem{Belov:2015ufh}
V.~Belov {\em et~al.}, ``{The vGeN experiment at the Kalinin Nuclear Power
  Plant},'' \href{http://dx.doi.org/10.1088/1748-0221/10/12/P12011}{{\em JINST}
  {\bfseries 10} (2015) P12011}.

\bibitem{conus}
Private communication with conus collaboration.

\bibitem{Billard:2016giu}
J.~Billard {\em et~al.}, ``{Coherent Neutrino Scattering with Low Temperature
  Bolometers at Chooz Reactor Complex},''
  \href{http://dx.doi.org/10.1088/1361-6471/aa83d0}{{\em J.Phys.} {\bfseries
  G44} (2017) 105101}, \href{http://arxiv.org/abs/1612.09035}{{\ttfamily
  arXiv:1612.09035 [physics.ins-det]}}.

\bibitem{Strauss:2017cuu}
R.~Strauss {\em et~al.}, ``{The $\nu$-cleus experiment: A gram-scale
  fiducial-volume cryogenic detector for the first detection of coherent
  neutrino-nucleus scattering},''
  \href{http://dx.doi.org/10.1140/epjc/s10052-017-5068-2}{{\em Eur.Phys.J.}
  {\bfseries C77} (2017) 506},
  \href{http://arxiv.org/abs/1704.04320}{{\ttfamily arXiv:1704.04320
  [physics.ins-det]}}.

\bibitem{Fricke:1995zz}
G.~Fricke {\em et~al.}, ``{Nuclear Ground State Charge Radii from
  Electromagnetic Interactions},''
  \href{http://dx.doi.org/10.1006/adnd.1995.1007}{{\em Atom.Data Nucl.Data
  Tabl.} {\bfseries 60} (1995) 177--285}.

\bibitem{Angeli:2013epw}
I.~Angeli and K.~Marinova, ``{Table of experimental nuclear ground state charge
  radii: An update},'' \href{http://dx.doi.org/10.1016/j.adt.2011.12.006}{{\em
  Atom.Data Nucl.Data Tabl.} {\bfseries 99} (2013) 69--95}.

\bibitem{Abrahamyan:2012gp}
S.~Abrahamyan {\em et~al.}, ``{Measurement of the Neutron Radius of 208Pb
  Through Parity-Violation in Electron Scattering},''
  \href{http://dx.doi.org/10.1103/PhysRevLett.108.112502}{{\em Phys.Rev.Lett.}
  {\bfseries 108} (2012) 112502},
  \href{http://arxiv.org/abs/1201.2568}{{\ttfamily arXiv:1201.2568 [nucl-ex]}}.

\bibitem{Horowitz:2012tj}
C.~Horowitz {\em et~al.}, ``{Weak charge form factor and radius of 208Pb
  through parity violation in electron scattering},''
  \href{http://dx.doi.org/10.1103/PhysRevC.85.032501}{{\em Phys.Rev.}
  {\bfseries C85} (2012) 032501},
  \href{http://arxiv.org/abs/1202.1468}{{\ttfamily arXiv:1202.1468 [nucl-ex]}}.

\bibitem{Orrigo:2016wgu}
S.~Orrigo, L.~Alvarez-Ruso, and C.~Pe{\~n}a-Garay,
  \href{http://dx.doi.org/10.1016/j.nuclphysbps.2015.09.060}{``{A New Approach
  to Nuclear Form Factors for Direct Dark Matter Searches},''} vol.~273-275,
  pp.~414--418.
\newblock 2016.

\bibitem{Cadeddu:2018izq}
M.~Cadeddu and F.~Dordei, ``{Reinterpreting the weak mixing angle from atomic
  parity violation in view of the Cs neutron rms radius measurement from
  COHERENT},'' \href{http://dx.doi.org/10.1103/PhysRevD.99.033010}{{\em
  Phys.Rev.} {\bfseries D99} (2019) 033010},
  \href{http://arxiv.org/abs/1808.10202}{{\ttfamily arXiv:1808.10202
  [hep-ph]}}.

\bibitem{Huang:2019ene}
X.-R. Huang and L.-W. Chen, ``{Neutron Skin in CsI and Low-Energy Effective
  Weak Mixing Angle from COHERENT Data},''
  \href{http://arxiv.org/abs/1902.07625}{{\ttfamily arXiv:1902.07625
  [hep-ph]}}.

\bibitem{AristizabalSierra:2019zmy}
D.~Aristizabal~Sierra, J.~Liao, and D.~Marfatia, ``{Impact of form factor
  uncertainties on interpretations of coherent elastic neutrino-nucleus
  scattering data},'' \href{http://dx.doi.org/10.1007/JHEP06(2019)141}{{\em
  JHEP} {\bfseries 1906} (2019) 141},
  \href{http://arxiv.org/abs/1902.07398}{{\ttfamily arXiv:1902.07398
  [hep-ph]}}.

\bibitem{Lewin:1995rx}
J.~Lewin and P.~Smith, ``{Review of mathematics, numerical factors, and
  corrections for dark matter experiments based on elastic nuclear recoil},''
  \href{http://dx.doi.org/10.1016/S0927-6505(96)00047-3}{{\em Astropart.Phys.}
  {\bfseries 6} (1996) 87--112}.

\bibitem{Helm:1956zz}
R.~H. Helm, ``{Inelastic and Elastic Scattering of 187-Mev Electrons from
  Selected Even-Even Nuclei},''
  \href{http://dx.doi.org/10.1103/PhysRev.104.1466}{{\em Phys.Rev.} {\bfseries
  104} (1956) 1466--1475}.

\bibitem{Sprung_1997}
D.~W.~L. Sprung and J.~Martorell, ``The symmetrized fermi function and its
  transforms,'' \href{http://dx.doi.org/10.1088/0305-4470/30/18/026}{{\em
  Journal of Physics A: Mathematical and General} {\bfseries 30} no.~18, (Sep,
  1997) 6525--6534}. \url{https://doi.org/10.1088%2F0305-4470%2F30%2F18%2F026}.

\bibitem{Klein:1999qj}
S.~Klein and J.~Nystrand, ``{Exclusive vector meson production in relativistic
  heavy ion collisions},''
  \href{http://dx.doi.org/10.1103/PhysRevC.60.014903}{{\em Phys.Rev.}
  {\bfseries C60} (1999) 014903}.

\bibitem{Piekarewicz:2016vbn}
J.~Piekarewicz, A.~Linero, P.~Giuliani, and E.~Chicken, ``{Power of two:
  Assessing the impact of a second measurement of the weak-charge form factor
  of $^{208}$Pb},'' \href{http://dx.doi.org/10.1103/PhysRevC.94.034316}{{\em
  Phys.Rev.} {\bfseries C94} (2016) 034316},
  \href{http://arxiv.org/abs/1604.07799}{{\ttfamily arXiv:1604.07799
  [nucl-th]}}.

\bibitem{nndc}
\url{http://www.nndc.bnl.gov/ensdf}.

\bibitem{Beringer:1900zz}
{\bfseries Particle Data Group} Collaboration, J.~Beringer {\em et~al.},
  ``{Review of Particle Physics (RPP)},''
  \href{http://dx.doi.org/10.1103/PhysRevD.86.010001}{{\em Phys.Rev.}
  {\bfseries D86} (2012) 010001}.

\bibitem{Suhonen:2017rjf}
J.~Suhonen, ``{Impact of the quenching of $g_{\rm A}$ on the sensitivity of
  $0\nu\beta\beta$ experiments},''
  \href{http://dx.doi.org/10.1103/PhysRevC.96.055501}{{\em Phys. Rev.}
  {\bfseries C96} no.~5, (2017) 055501},
\href{http://arxiv.org/abs/1708.09604}{{\ttfamily arXiv:1708.09604 [nucl-th]}}.

\bibitem{Collar:2014lya}
J.~Collar {\em et~al.}, ``{Coherent neutrino-nucleus scattering detection with
  a CsI[Na] scintillator at the SNS spallation source},''
  \href{http://dx.doi.org/10.1016/j.nima.2014.11.037}{{\em Nucl.Instrum.Meth.}
  {\bfseries A773} (2015) 56--65},
  \href{http://arxiv.org/abs/1407.7524}{{\ttfamily arXiv:1407.7524
  [physics.ins-det]}}.

\bibitem{Kortelainen:2006rd}
M.~Kortelainen, J.~Suhonen, J.~Toivanen, and T.~Kosmas, ``{Event rates for CDM
  detectors from large-scale shell-model calculations},''
  \href{http://dx.doi.org/10.1016/j.physletb.2005.10.057}{{\em Phys.Lett.}
  {\bfseries B632} (2006) 226--232}.

\bibitem{Toivanen:2009zza}
P.~Toivanen, M.~Kortelainen, J.~Suhonen, and J.~Toivanen, ``{Large-scale
  shell-model calculations of elastic and inelastic scattering rates of
  lightest supersymmetric particles (LSP) on I-127, Xe-129, Xe-131, and Cs-133
  nuclei},'' \href{http://dx.doi.org/10.1103/PhysRevC.79.044302}{{\em
  Phys.Rev.} {\bfseries C79} (2009) 044302}.

\bibitem{Papoulias:2013gha}
D.~Papoulias and T.~Kosmas, ``{Nuclear aspects of neutral current non-standard
  $\nu$-nucleus reactions and the role of the exotic $\mu^-\to e^{-}$
  transitions experimental limits},''
  \href{http://dx.doi.org/10.1016/j.physletb.2013.12.028}{{\em Phys.Lett.}
  {\bfseries B728} (2014) 482--488},
  \href{http://arxiv.org/abs/1312.2460}{{\ttfamily arXiv:1312.2460 [nucl-th]}}.

\bibitem{Coraggio:2017bqn}
L.~Coraggio, L.~De~Angelis, T.~Fukui, A.~Gargano, and N.~Itaco, ``{Calculation
  of Gamow-Teller and two-neutrino double- {\ensuremath{\beta}} decay
  properties for $^{130}$Te and $^{136}$Xe with a realistic nucleon-nucleon
  potential},'' \href{http://dx.doi.org/10.1103/PhysRevC.95.064324}{{\em
  Phys.Rev.} {\bfseries C95} (2017) 064324},
  \href{http://arxiv.org/abs/1703.05087}{{\ttfamily arXiv:1703.05087
  [nucl-th]}}.

\end{thebibliography}\endgroup

\end{document}